\documentclass[prl,a4paper,superscriptaddress,nofootinbib,twocolumn]{revtex4-1}

\usepackage{graphicx}
\usepackage{color}
\usepackage{amssymb,amsmath,amsfonts}
\usepackage{latexsym}
\usepackage{array,tabularx}
\usepackage{dcolumn}
\usepackage{enumerate}
\usepackage{hyperref}
\usepackage{mathtools}
\usepackage{float}
\usepackage{xcolor}
\usepackage[T1]{fontenc}

\newcommand{\bs}[1]{ {\boldsymbol{#1}} }

\newcommand{\pdd}[2]{\frac{\partial{#1}}{\partial{#2}}}

\begin{document}

\title{First-order dissipative phase transition in an exciton-polariton condensate}

\author{Galbadrakh~Dagvadorj}
\affiliation{Department of Physics and Astronomy, University College London, Gower Street, London, WC1E 6BT, UK}
\affiliation{Department of Physics, University of Warwick, Coventry, CV4 7AL, UK}

\author{Micha\l{} Kulczykowski}
\affiliation{Institute of Physics, Polish Academy of Sciences,\\ Aleja Lotnik\'ow 32/46, PL-02-668 Warsaw, Poland}

\author{Marzena~H.~Szyma\'nska}
\email{m.szymanska@ucl.ac.uk}
\affiliation{Department of Physics and Astronomy, University College London, Gower Street, London, WC1E 6BT, UK}

\author{Micha\l{} Matuszewski}
\email{mmatu@ifpan.edu.pl}
\affiliation{Institute of Physics, Polish Academy of Sciences,\\ Aleja Lotnik\'ow 32/46, PL-02-668 Warsaw, Poland}

\begin{abstract}

We investigate the phase diagram of a two-dimensional driven-dissipative system of polaritons coupled to the excitonic reservoir. We find that two critical points exists. The first corresponds to the quasi-condensation and the second to a first-order phase transition from the non-uniform state with spatially modulated density to a uniform state. The latter is related to the modulational instability of a homogeneous state due to the repulsive interactions with the non-condensed reservoir. The first-order character of the transition is evidenced by a discontinuity in the density and the correlation length as well as the phase coexistence and metastability. Moreover, we show that a signature of a Berezinskii-Kosterlitz-Thouless-like transition can be observed in the non-uniform phase.

\end{abstract} 

\maketitle

Some non-equilibrium phase transitions are shown to be universal and occur across different time and length scales: from active biological matter~\cite{Cugliandolo_2017,Cugliandolo_FullPhaseDiagram} and self-organized systems~\cite{Nicolis_SelfOrganization} to quantum fluids of light~\cite{Sieberer_DynamicalCritical,Zamora_Criticalities,Hanai}.  Non-equilibrium phases can be characterised by exotic phenomena with no counter-parts in equilibrium settings, for example the coexistence of first-order and Berezinskii-Kosterlitz-Thouless phase transitions~\cite{Cugliandolo_2017,Cugliandolo_FullPhaseDiagram,Domany_FirstOrder,Enter_FirstOrder}. Active matter and classical dissipative systems lead to a plethora of such behaviours. However, at the quantum level the possibilities to investigate exotic non-equilibrium phase transitions are more limited.

Exciton-polaritons are quantum quasiparticles that result from the strong coupling of light and matter in semiconductor microcavities~\cite{Kavokin_Microcavities,Hopfield_Polaritons,Weisbuch_Polaritons}.
Polaritons provide a level of complexity of nonequilibrium physics that is comparable to biological systems, while keeping the quantum character. Furthermore, coupling to the additional degrees of freedom of the reservoir provides a unique  opportunity to study phase transitions in the quantum domain, but in contact with a classical environment.
The distinctive properties of polaritons include ultralow effective mass, picosecond time-scales, and strong interactions. At the same time, polaritons can be directly observed by optical methods due to the coupling to the outside of the microcavity. These remarkable features made it possible to realise with polaritons a number of fascinating phenomena, including bosonic condensation~\cite{Kasprzak_BEC}, formation of vortices~\cite{Vina_VorticesCoherently,Deveaud_VortexDynamics} and superfluid-like states~\cite{Amo_Superfluidity,Marzena_Superfluidity}.

Due to their driven-dissipative nature resulting from the finite lifetime, polaritons are particularly well suited to the investigation of dissipative phase transitions. Such transitions occur in systems far from thermal equilibrium and do not correspond to standard universality classes.  In driven-dissipative systems, external pumping plays the role of temperature. Recently, particular attention was devoted to universal properties and classification of non-thermal critical points~\cite{Sieberer_DynamicalCritical,Altman_DrivenSuperfluid2D,Wouters_SpatialCoherence,Zamora_Criticalities}, Berezinskii-Kosterlitz-Thouless (BKT) transitions~\cite{Szymanska_NonequilibriumBKT,Sanvitto_TopologicalOrder}, the Kibble-\.Zurek mechanism of defect formation~\cite{Matuszewski_UniversalityPolaritons,Malpuech_KZM,Zamora_KZM}, phase ordering kinetics following sudden quenches~\cite{Kulczykowski_PhaseOrdering,Comaron_DynamicalCriticalExponents} and critical dynamics~\cite{Bobrovska_CriticalDynamicsTreeLike}. However, most of these studies investigated continuous phase transition, which in the polariton context can be related to the quasi-condensation critical point.A first-order dissipation-driven phase transition with an endpoint was predicted in a uniform condensate~\cite{Hanai}. Recently, it was demonstrated both theoretically~\cite{Wouters_Excitations,Bobrovska_Stability,Liew_InstabilityInduced} and experimentally~\cite{Bobrovska_DynamicalInstability,Estrecho_SingleShotCondensation,Bloch_UnstableRegimes} that another critical point, corresponding to dynamical (modulational) instability, exists in polariton systems, but its critical properties remained unknown.

Here, we explore in detail the phase diagram of a general model of polariton condensation, based on the Truncated Wigner approach, including the interaction with the non-condensed reservoir. In particular, we investigate the properties of the system close to criticality. We find that the second critical point corresponds to a dissipative phase transition of the \emph{first-order} type. The transition is from the non-uniform spatially modulated state to the state with a homogeneous density. We observe the most striking characteristics of a first-order phase transition, including sudden jumps of the average density and correlation length in the absence of scaling laws. Moreover, we find the region of metastability in the parameter space, which is another important property associated with a first-order phase transition~\cite{Binder_FirstOrderPTs}. In a dynamical study, the coexistence of phases is revealed in a transient state at parameters close to the critical point. Interestingly, the phase diagram of the system is analogous to the one of the Vicsek model of flocking of self-propelled particles (e.g. flying birds, schools of fish, herds of quadrupeds, bacterial colonies among others), where the transition from the disordered to the ordered particle directions (the condensate phase in our model) is accompanied by a region in parameter space where pattern formation emerges~\cite{Gregoire_Flocking,Mishra_Flocking}. As in the present case, the discontinuous character of the corresponding transition is due to the coupling to the particle density field~\cite{Mishra_Flocking,Bobrovska_Adiabatic}.

Finally, we explore the possibility of coexistence of the first-order  and the BKT-like phase transitions. Previously, such coexistence has been found in classical XY models~\cite{Domany_FirstOrder,Enter_FirstOrder} and active particle systems~\cite{Cugliandolo_2017} with a transition from the exponential decay of correlations, to algebraic decay and finally to long-range order. This highlights the importance of considering the reservoir, since the models of polariton condensation which eliminate this degree of freedom do not manifest such coexistence~\cite{Sieberer_DynamicalCritical,Szymanska_NonequilibriumBKT}. We find that the vortex-antivortex pairing, which is fundamental for BKT physics, occurs in the non-uniform state slightly above the mean-field condensation threshold. At the same time, the direct measurement of the correlation function does not reveal BKT scaling laws due to the breakdown of spatial homogeneity of the system. The non-uniform state turns out to consist of mutually uncorrelated condensate islands, which limits the maximum correlation length to the size of a single island.

\paragraph*{System and the Model.}

The stochastic equation for the lower polariton (LP) field  $\psi(\bs{r},t)$ within the Truncated Wigner approach \cite{Wouters_ClassicalFields},
\begin{align}
\label{eq:GPE}
  i \hbar\, d\psi &= \bigg [ \frac{-\hbar^2}{2m_\textsc{lp}} \nabla_\bs{r}^2+ g |\psi|^2_-+ g_\textsc{r} n_\textsc{r}  
  +\frac{i\hbar}{2} \left( R_{\rm sc} n_\textsc{r} -\gamma \right) \bigg] \psi d t \nonumber \\
  &+  i \hbar \sqrt{\frac{\hbar\left( R_{\rm sc} n_\textsc{r} +\gamma \right)}{4\Delta V}}d W(\bs{r},t),
  \end{align}
is coupled with the reservoir density $n_\textsc{r}(\bs{r},t)$, 
 \begin{equation*} 
\pdd{n_\textsc{r}}{t} = P(\bs{r}) - \gamma_\textsc{r} n_\textsc{r}-R_{\rm sc} n_\textsc{r} |\psi|^2_-, 
\end{equation*}
where 
$m_\textsc{lp}$ is the effective mass of lower polaritons,
$g$ the effective inter-polariton interaction strength,
$R_{\rm sc} n_\textsc{r}$  the rate of scattering from the reservoir to the polariton condensate,
and $\gamma$ the decay rate of lower polaritons.
$P$ is the pump rate, i.e. the rate at which excitons are added to the reservoir,
$\gamma_\textsc{r}$ the loss rate of the reservoir while
$g_\textsc{r}$ is the strength of the interaction between polaritons and particles in the reservoir.
$d W$ is the derivative of a Wiener process, corresponding to quantum noise, and
$\Delta V$ the volume of a unit cell in the simulation.
In the above equation, the term $|\psi(\bs{r},t)|^2_- \equiv |\psi(\bs{r},t)|^2 -\frac{1}{\Delta V}$ takes into account the quantum noise correction in the evaluation of the polariton density.


In the mean field limit, the noise is absent in the equations of motion, and $|\psi(\bs{r},t)|^2_-$ is replaced by $|\psi(\bs{r},t)|^2$. The system can be treated analytically under the assumption of a homogeneous steady state, $\psi(\bs{r},t)=\psi_0 {\rm e}^{-i\mu t}$ and $n_\textsc{r}(\bs{r},t)=n_\textsc{r}^0$. The solution depends on whether the pump power $P$ is higher than the condensation threshold given by $P_{\rm th}^\text{MF}=\gamma_\textsc{r} n_\textsc{R}$. Below threshold, the polariton density is zero and the reservoir density is $n_\textsc{r}^0=P/\gamma_\textsc{r}$, and above threshold the polariton  density is nonzero $|\psi|^2=(P-P_{\rm th}^\text{MF})/\gamma$ while $n_\textsc{r}=\gamma/R_\textsc{Sc}$.

The second critical point is related to the stability of the homogeneous solution. As it turns out, the homogeneous quasi-condensate may not be stable at low pump powers and is replaced by a spatially modulated, non-uniform state with condensate islands separated by low density regions of normal state polaritons. This dynamical instability of polariton condensates was studied in detail in~\cite{Wouters_Excitations,Bobrovska_Stability,Liew_InstabilityInduced} and its relation to the breakdown of the adiabatic approximation was analysed in~\cite{Bobrovska_Adiabatic}. The analytical condition for stability of the homogeneous solution was derived in~\cite{Ostrovskaya_DarkSolitons} 
\begin{equation} \label{eq:stability}
\frac{P}{P_\text{MF}}>\frac{P_{\rm MI}}{P_\text{MF}}=\frac{g_\textsc{r}}{g}\frac{\gamma}{\gamma_\textsc{r}}.
\end{equation}
Recently, the instability was demonstrated experimentally in organic and inorganic polariton systems, both in the case of pulsed~\cite{Bobrovska_DynamicalInstability,Estrecho_SingleShotCondensation} and continuous pumping~\cite{Bloch_UnstableRegimes}. In particular, it was shown that the instability leads to a dramatic modulation of the polariton density and splitting of the condensate into separate droplets, analogous to the spatial hole burning effect in lasers~\cite{Bobrovska_DynamicalInstability,Estrecho_SingleShotCondensation}.

Polariton systems described by Eqs.~(\ref{eq:GPE}) and similar models have been studied in the context of critical behaviour~\cite{Sieberer_DynamicalCritical,Comaron_DynamicalCriticalExponents,Szymanska_NonequilibriumBKT,Sanvitto_TopologicalOrder,Kulczykowski_PhaseOrdering,Matuszewski_UniversalityPolaritons,Malpuech_KZM}. However, to date the theoretical studies were concerned with the first critical point, which corresponds to the quasi-condensation phase transition. At the same time, very little is known about the critical properties of the second critical point~\cite{Bobrovska_CriticalDynamicsTreeLike}, corresponding to the transition between homogeneous and spatially non-uniform condensates. In this work, we investigate in detail the phase diagram of the model, in particular focusing on the critical behaviour around the second critical point given by Eq.~(\ref{eq:stability}). We demonstrate striking similarities to the physics of equilibrium first-order phase transitions, despite the fact that we consider a driven-dissipative system. In particular, we show discontinuous behaviour of the correlation function, phase separation, and multistability of steady state solutions, all being considered evidence of a first-order phase transition~\cite{Binder_FirstOrderPTs}.

\begin{figure}
\begin{center}
\includegraphics[width=0.5\textwidth]{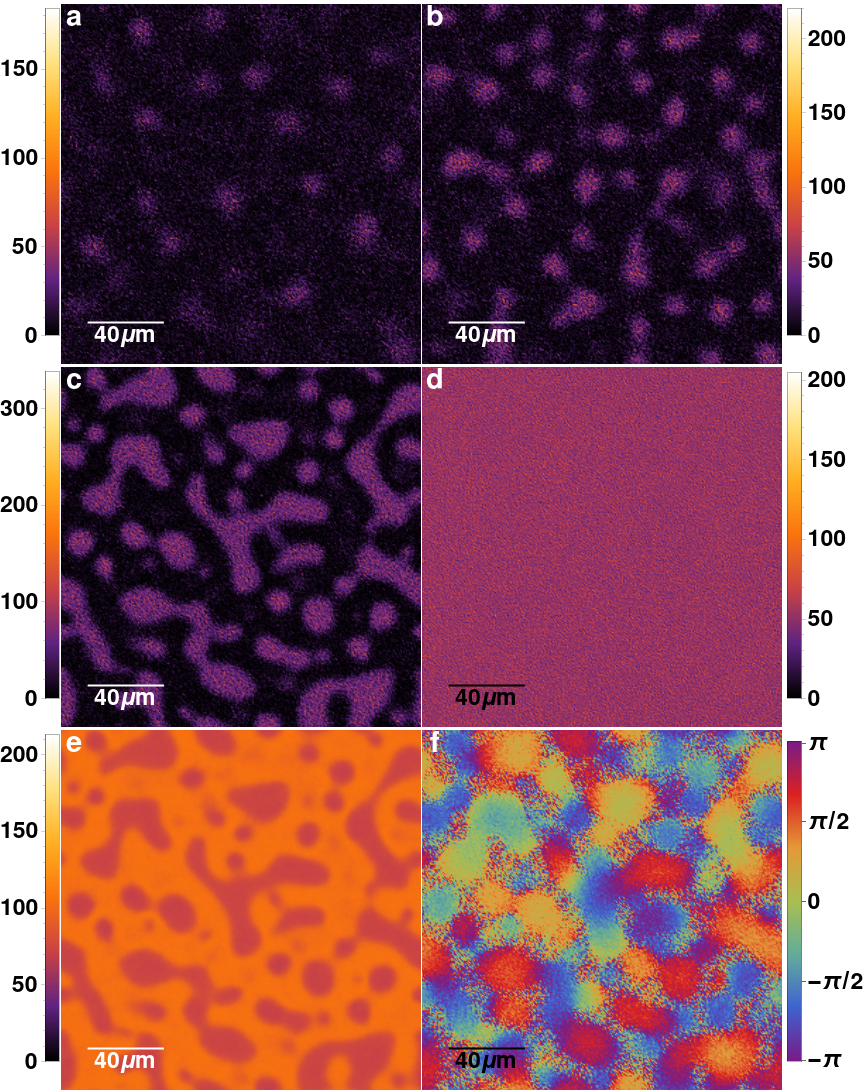}
\end{center}
\caption{ Snapshots of the steady-state polariton density profiles $|\psi|^2$ for increasing pump power (a)-(d). 
	 The spatially non-uniform states for $P=1.01P_{th}^{MF}$, 
	$P=1.1P_{th}^{MF}$ and $P=1.3P_{th}^{MF}$ are shown in panels (a)-(c) while the homogeneous state at $P=1.4P_{th}^{MF}>P_{MI}=1.(3)P_{th}^{MF}$ in panel (d). 
	Reservoir density $n_R$ (e) and the condensate phase (f) correspond to panel (c). All density color scales are in $(\mu$m$^{-2})$. }\label{fig:snapshots}
\end{figure}

We begin by investigating the steady state of the system obtained after a long evolution at constant homogeneous pumping in a box with periodic boundary conditions. We start each simulation from the mean-field steady state solution. The system is evolved for a long time in each case to make sure that it has reached the steady state. The typical results for realistic values of physical parameters are shown in Fig.~\ref{fig:snapshots}. The parameters are taken from~\cite{Yamamoto_VortexPair} and are typical for average semiconductor microcavities, i.e.~$m_\textsc{lp} = 5\times10^{-5}m_{\text{e}}$, $\tau = 3\,$ps, $\tau_\textsc{r} = 2\,$ps, $g = 6.0\times10^{-3}\,$meV$\mu$m$^2$, $g_\textsc{r} = 12.0\times10^{-3}\,$meV$\mu$m$^2$, $\tau = 2.0\,$ps, while $R_{\rm sc} = 0.001\,\mu$m$^2$ps$^{-1}$.

Figs.~\ref{fig:snapshots}(a)-(d) show the steady-state density profiles of the polariton field for increasing external pumping, where brighter regions correspond to higher density. For the pump power slightly above threshold in Fig.~\ref{fig:snapshots}(a), a sparse set of islands with higher  density is formed, which correspond to localised condensation centres. The positions of the islands are random and vary from one realisation of the noise to another. Between the islands the density is very low and dominated by  noise. When increasing pumping, the highly populated areas extend, and the islands start to merge as in Figs.~\ref{fig:snapshots}(b,c). The average polariton density also grows as indicated by the change of the color scale. Finally, when the pumping is increased above the stability threshold~(\ref{eq:stability}), a homogeneous density profile with a slight addition of noise is obtained, see Fig.~\ref{fig:snapshots}(d). We call this state ``uniform'' in contrast to the ``non-uniform'' states in Figs.~\ref{fig:snapshots}(a)-(c). This behaviour is in agreement with previous results \cite{Bobrovska_Stability,Ostrovskaya_DarkSolitons,Bobrovska_DynamicalInstability,Estrecho_SingleShotCondensation}.  Below the second threshold, the areas of low polariton density correspond to higher reservoir density, as shown in Fig.~\ref{fig:snapshots}(e). It is important to note that the phase of the polariton field, shown in Fig.~\ref{fig:snapshots}(f), suggests that while each of the polariton density islands has approximately constant phase, there is no phase coherence between different islands.  This partial coherence bears similarity to the Griffiths phase in dilute Ising ferromagnets, where partial order is maintained in each of mutually disconnected islands~\cite{Griffiths_phase}, and the discontinuous transition may be related to the divergence of susceptibility in the Griffiths phase. In this sense, the state of the whole system below the stability threshold can be related to the phenomenon of a kind of quasi-condensation, distinct from the well-known homogeneous 2D quasi-condensation below the BKT transition which is characterised by algebraic decay of coherence over the whole system size \cite{Szymanska_NonequilibriumBKT}.

\begin{figure}
\begin{center}
\includegraphics[width=0.5\textwidth]{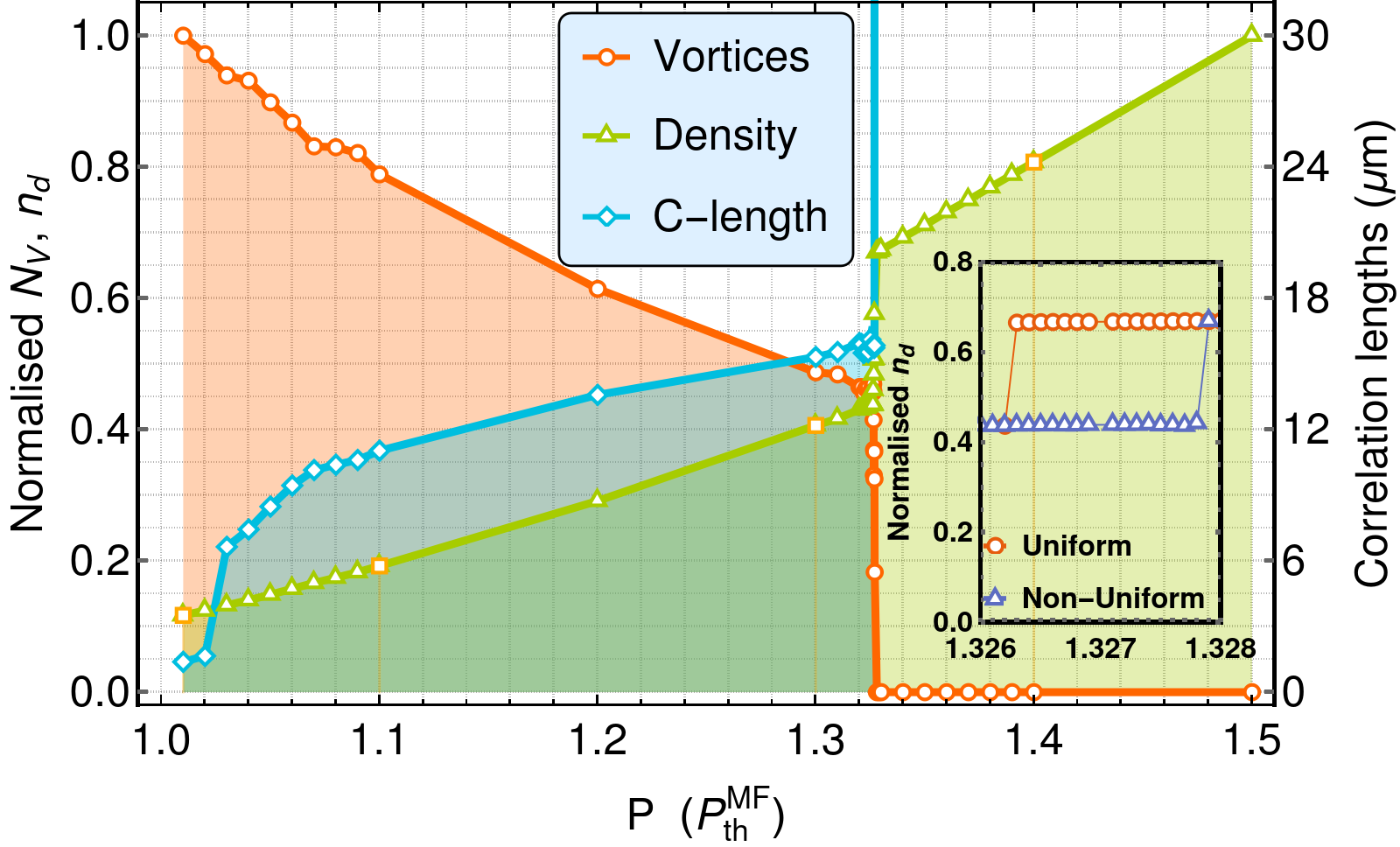}
\end{center}
\caption{Phase diagram with the first-order transition from the spatially modulated state at
        low pump powers to the homogeneous state at high powers. The transition is marked by the discontinuity
        in the density (green triangles) and the sudden disappearance of vortices (orange circles). The correlation length (cyan diamonds) as a
        function of pump power reveals two transitions: the second-order condensation transition
        at $P=P_{\rm th}$, and the modulational instability transition at $P=P_{MI}\approx 1.33 P^{\rm MF}_{\rm th}$, and is  
        of the order of the system size (beyond scale) in the homogeneous phase with $P>P_{MI}$. Note, that
        the condensation threshold $P_{\rm th}$ is slightly shifted with respect to its mean-field value, $P_{th}^{MF}$.
	The cases shown in Fig.~\ref{fig:snapshots} are marked with orange open squares.}\label{fig:phasediagram}
\end{figure}

\paragraph*{First order phase transition.}
In Fig.~\ref{fig:phasediagram} we show the phase diagram of the system, beginning from the homogeneous quasi-condensation threshold $P_{\rm th}$. Note that below $P_{\rm th}$, the polariton density is very low and the system is dominated by fluctuations. Above threshold, the average density is gradually increasing, in qualitative agreement with the mean-field analytical predictions. In Fig.~\ref{fig:phasediagram} we also plot the first-order (phase) correlation length $\bs{\xi}_c$, calculated from the formula $g(\bs{\xi}_c,t)^{(1)}=e^{-1}$ where 
$$g(\bs{r},t)^{(1)}=\frac{\langle\psi^*(\bs{r}+\bs{u},t)\psi(\bs{u},t)\rangle-\delta_{\bs{r}+\bs{u},\bs{u}}/2dV}{\langle|\psi^*(\bs{r}+\bs{u},t)|^2\rangle \langle|\psi(\bs{u},t)|^2\rangle},$$ $\langle ... \rangle$ denotes averaging over both noise realisations and the auxiliary position $\bs{u}$, and the number of vortices, obtained  using the algorithm of vortex core counting~\cite{Szymanska_NonequilibriumBKT}.  The above formula results from the general correspondence rules between averages of quantum operators and stochastic fields in the truncated Wigner approximation. These rules follow from the derivation of stochastic process from the Master equation for the density matrix of the open system~\cite{Wouters_ClassicalFields,Polkovnikov_PhaseSpace}. In particular, the correction term $-\delta_{\bs{r}+\bs{u},\bs{u}}/2dV$ arises from the symmetric ordering of operators in the truncated Wigner approximation, which is different from normal ordering. The actual condensation threshold $P_{\rm th}$ is slightly shifted with respect to its mean-field value  $P_{\rm th}^{\rm MF}$. The correlation length increases quickly above threshold, which is related to the high coherence of the condensation islands of Fig.~\ref{fig:snapshots}(a)-(c). The number of vortices is decreasing, which is 
also an  indicator of the increase in coherence.

At the second threshold $P_{MI}$, a dramatic change in the state of the system can be noticed. There is a sudden jump in the average density, the correlation length, and the number of vortices. The correlation length becomes comparable to the size of the system, and the number of vortices drops to zero, which indicates the appearance of a fully condensed state as in Fig.~\ref{fig:snapshots}(d) with a uniform density and slowly varying phase throughout the system. This is in striking contrast to continuous (higher-order) phase transitions, where correlation length tends to infinity in a continuous manner at the critical point, following an appropriate scaling law~\cite{Szymanska_NonequilibriumBKT,Sieberer_DynamicalCritical}, and there is no discontinuity of the density.

Further evidence of the first-order transition can be obtained by a closer inspection of the system in the vicinity of the critical point. The inset in Fig.~\ref{fig:phasediagram} shows the density as a function of pump power in the small area around the threshold. 
We find that for a certain range of pump powers, the steady state can be either homogeneous  or inhomogeneous at the same value of pumping for different noise realisations. The circles and triangles in the inset show the average density in these two types of states. This behaviour resembles a standard first-order phase transition, where metastability of phases is present in a certain range of parameters close to the critical point~\cite{Binder_FirstOrderPTs}. 

\begin{figure}
\begin{center}
\includegraphics[width=0.5\textwidth]{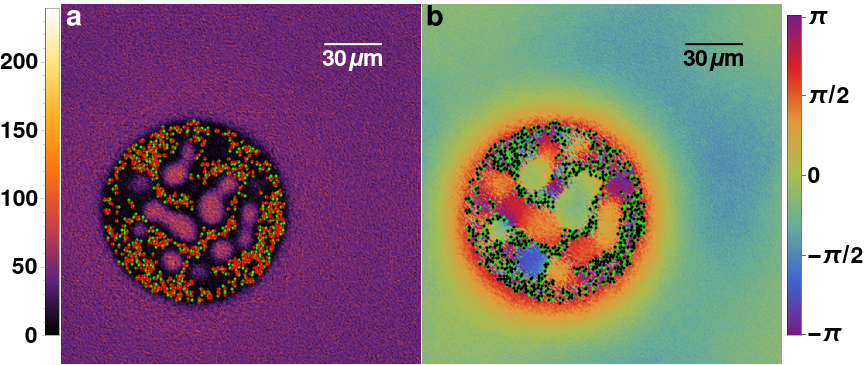}
\end{center}
\caption{Separation between homogeneous and spatially modulated phases in a transient state during the evolution from a mean-field homogeneous state at $P=1.326P_{th}^{MF}$. The density (a) and  the phase  (b) of the field, with red and green dots marking  vortices and antivortices in low-density regions.}\label{fig:phaseseparation}
\end{figure}

In the critical region, another characteristic of a first-order transition, i.e.~phase separation and the coexistence of phases, can be observed. In Fig.~\ref{fig:phaseseparation} we show a snapshot of the polariton density and phase of a state emerging during the evolution from the initial noise at the pump power $P=1.326P_{th}^{MF}$. In Fig.~\ref{fig:phaseseparation} the positions of vortices, indicated by dots, are superimposed on the polariton density (a) and phase (b) profiles. The system clearly develops two distinct regions, the inner one containing the non-uniform density and phase and the outer one containing the homogeneous density and smoothly varying phase. We find that this state is an unstable transient in an evolution leading finally to a non-uniform state in the whole computational area. Nevertheless, this clearly shows that the two phases can coexist in the system, although with an unstable boundary, which in this case eventually favours the non-uniform state. 

\begin{figure}
\begin{center}
	\includegraphics[width=0.47\textwidth]{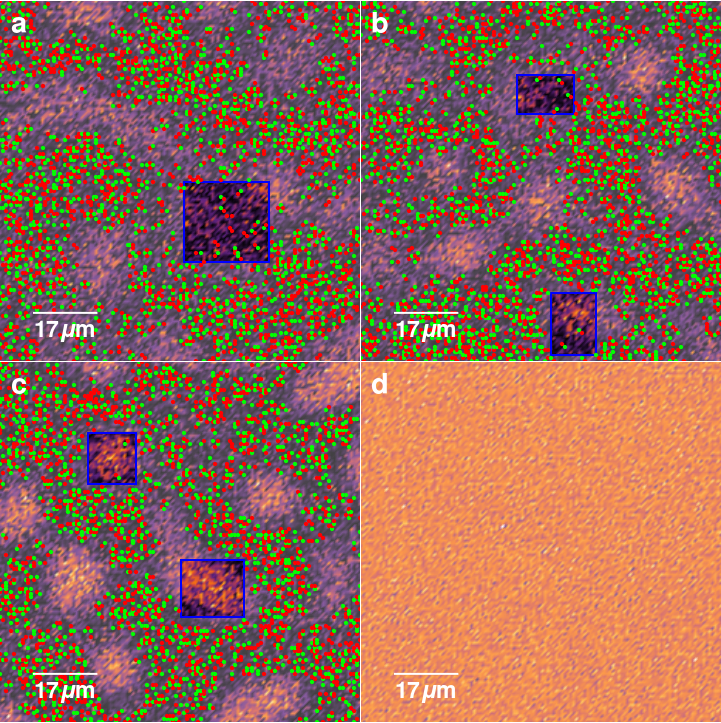}
\end{center}
\caption{Vortices (red) and anti-vortices (green) with super-imposed polariton density in the steady states at 
	pump powers (a) $P=1.01P_{th}^{MF}$, (b) $P=1.03P_{th}^{MF}$, (c) $P=1.09P_{th}^{MF}$ and (d) $P=1.327P_{th}^{MF}$. 
	The regions with overlapping dots are of low-density, where the condensate phase is 
	dominated by noise. In (a)-(c) we highlighted typical vortex configurations appearing in high-density regions. At low power (a), many unpaired vortices are visible. As the power is increased, pairing of vortices and antivortices occurs (b), and at an even higher power there are regions with no vortices and regions with only a few vortex-antivortex pairs (c). This pairing is an analog of the BKT transition in the 
	thermodynamic limit of a homogeneous system.} \label{fig:BKT}
\end{figure}

\paragraph*{Inhomogeneous  BKT-like transition.}
Finally, we investigate the possibility of detecting signatures of a BKT-like phase transition in the non-uniform phase. In Fig.~\ref{fig:BKT} we show examples of vortex and antivortex configurations superimposed on density profiles of the polariton field at four values of the external pumping. One of the important characteristics of the BKT transition is the pairing of vortices and antivortices close to the critical point. In our system, such pairing can be observed as the pumping is increased, which in a driven-dissipative polariton system is analogous to decreasing the temperature~\cite{Szymanska_NonequilibriumBKT}. In all panels of Fig.~\ref{fig:BKT} a large number of vortices and antivortices can be seen in the low density regions; however, these vortices result from the noise. A clear difference between the three cases can be noticed in the high density islands, which are not dominated by vortices. In Fig.~\ref{fig:BKT}(a), at very low pumping $P=1.01P_{th}^{MF}$ there are multiple unpaired vortices and antivortices in the high density islands. On increasing the pumping to $P=1.03P_{th}^{MF}$, almost all vortices and anti-vortices form pairs as visible in Fig.~\ref{fig:BKT}(b), which drastically increases the coherence length, see Fig.~\ref{fig:phasediagram}. At an even higher pumping $P=1.09P_{th}^{MF}$, practically all vortices disappear from the density islands, which become locally coherent quasicondensates, as shown in Fig.~\ref{fig:phasediagram}(c). Therefore, although the BKT-type algebraic decay of correlations cannot be observed in the non-uniform state due to the lack of coherence between islands, the system still displays an important signature of a BKT-like phase transition.

\paragraph*{Summary and Outlook.}
In conclusion, we have studied the phase diagram of a driven-dissipative system of non-resonantly pumped polaritons. We find that the second critical point exhibits characteristic features of a first-order  transition, despite the system being strongly out of equilibrium. The two critical points are supplemented by the BKT-like vortex binding transition, which can be studied in the non-uniform phase. Our results demonstrate the versatility of polaritons for the investigation of driven-dissipative phase transitions in a quantum system, and provide an analogy with previously studied classical systems. Interestingly, the first order and BKT transitions were previously found in a classical XY model with a specific form of the interaction potential~\cite{Domany_FirstOrder,Enter_FirstOrder}. This suggests that an analogous effective equation might be derived in the case of polaritons interacting with a reservoir, where it may appear as an effective equation for the phase. We expect that such a model, the derivation of which is beyond the scope of this work, would display rich behaviour and competition between the timescales of the condensate and the reservoir.

We thank Leticia Cugliandolo for stimulating discussions, and for making us aware of the presence of similar physics in active matter.
We acknowledge financial support from the Quantera ERA-NET cofund project InterPol (through the National Science Center, Poland, Grant No. 366 2017/25/Z/ST3/03032 (M.M.) and the EPSRC  Grant No. EP/R04399X/1 (M.H.S.)), EPSRC Grant No. EP/K003623/2 and EP/S019669/1 (M.H.S.), and from the National Science Centre, Poland, Grant No.~2016/22/E/ST3/00045 (M.K. and M.M.)

\bibliography{references}

\end{document}